\documentclass{optica-article}
\journal{optica}
\articletype{Research article}

\usepackage{subcaption}

\begin{document}

\title{Synthetic FM triplet for AM-free precision laser stabilization and spectroscopy}

\author{
Dhruv Kedar,\authormark{1,2,*}
Zhibin Yao,\authormark{1,2}
Ivan Ryger,\authormark{1}
John L. Hall,\authormark{1}
and Jun Ye\authormark{1}
}

\address{
\authormark{1}JILA, NIST and University of Colorado, 440 UCB, Boulder, Colorado 80309, USA \\
\authormark{2} Equal contributions
}

\email{\authormark{*}dhruv.kedar@colorado.edu}

\begin{abstract}
The Pound-Drever-Hall (PDH) cavity-locking scheme has found prevalent uses in precision optical interferometry and laser frequency stabilization. A form of frequency modulation spectroscopy, PDH enjoys superior signal-to-noise recovery, large acquisition dynamic range, wide servo bandwidth, and robust rejection of spurious effects. However, residual amplitude modulation at the signal frequency, while significantly suppressed, still presents an important concern for further advancing the state-of-the-art performances. Here we present a simplified and improved scheme for PDH using an acousto-optic modulator to generate digital phase reference sidebands instead of the traditionally used electro-optic modulator approach. We demonstrate four key advantages: (1) the carrier and two modulation tones are individually synthesized and easily reconfigured, (2) robust and orthogonal control of the modulated optical field is applied directly to the amplitude and phase quadratures, (3) modulation synthesis, demodulation, and feedback are implemented in a self-contained and easily reproducible electronic unit, and (4) superior active and passive control of residual amplitude modulation is achieved, especially when the carrier power is vanishingly low. These distinct merits stimulate new ideas on how we optimally enact PDH for a wide range of applications.
\end{abstract}

\section{Introduction}
Precision optical interferometry builds on transferring the stability of a length reference to an optical frequency. Generally an optical oscillator is stabilized to the reference using the Pound-Drever-Hall (PDH) method~\cite{Drever1983,Pound1946,HallHils1989,Eric2001}. In this scheme, modulation sidebands around a carrier act as phase reference to quantify frequency deviations of the carrier relative to the center of the reference’s spectral resonance. The versatility of this method has allowed its powerful application in a wide range of optical interferometers, from tabletop experiments seeking to produce low noise oscillators~\cite{Salomon1988,Julsgaard2007,Kessler2012,Matei2017,Zhang2017} to km-scale gravitational wave interferometers~\cite{Voyager,aLIGO,KAGRA2015,VIRGO,ET}. While alternative methods have been developed to allow homodyne detection of the reference’s error signal~\cite{Hansch1980,Shaddock1999,Diorico2022}, PDH detection is by far the most commonly used and achieves the most robust performance as the carrier and phase reference sidebands propagate along the same optical path and share the same wavefront. 

A key feature of the PDH technique is that the heterodyne beatnote between the cavity leakage reference and the incident field is formed with two out-of-phase modulation sidebands. In the presence of the resonance, this frequency discrimination scheme converts a frequency modulated (FM) spectrum into an amplitude modulated field, offering detection sensitivity limited only by the photon shot noise. Away from the cavity resonance, PDH should ideally produce a null baseline signal. However residual amplitude modulation (RAM), arising from any optical components in the beam path, can produce a fluctuating baseline.  Reduction of the spurious RAM has therefore been a desirable goal in FM spectroscopy, including the development of ultrastable optical interferometers/oscillators~\cite{Bjorklund1983,Whittaker1988,Kasapi2000,Kokeyama2014}. 

\begin{figure*}[t!]
    \centering
    \includegraphics[width=0.9\linewidth]{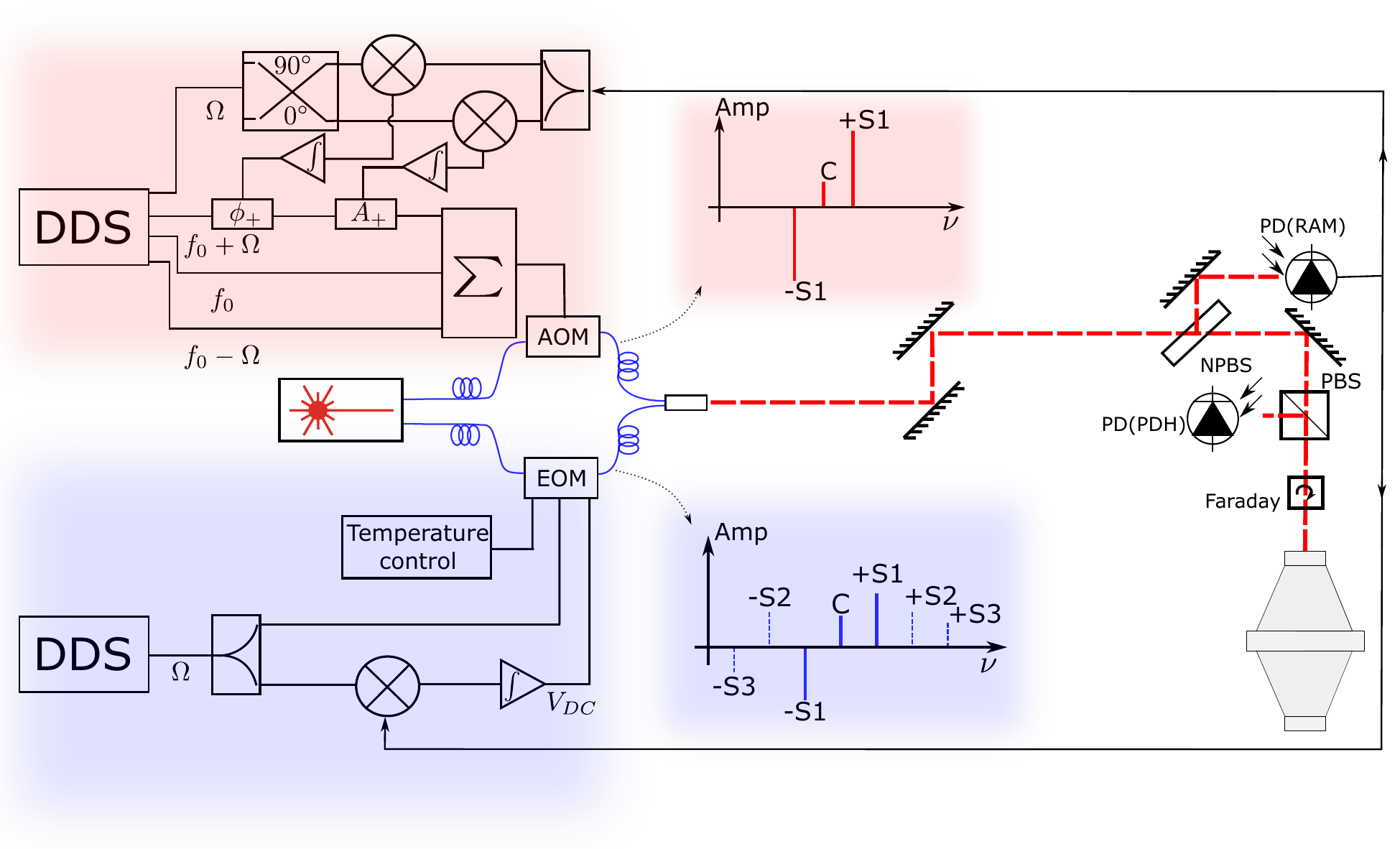}
    \caption{Optical and electronic experimental schema for AOM and EOM modulation configurations. The black, blue, and red dashed lines denote the electronic, fiber, and free-space optical connections, respectively. $\phi_+$: voltage controlled phase shifter; $A_+$: voltage controlled attenuator; $\Sigma$: RF power combiner; $V_{DC}$: Bias DC voltage applied on EOM for active RAM cancellation; PD: photo detector; NPBS: non polarizing beam splitter; PBS: polarizing beam splitter. Either the AOM or EOM configuration is used for PDH-based laser frequency stabilization.}
    \label{fig:schematic_opt}
\end{figure*}

In a typical PDH realization, an electro-optic modulator (EOM) is driven with a radio frequency (RF) signal $\Omega$ to produce a series of phase modulation sidebands around the incident optical carrier,
\begin{align}
    E e^{i(\omega t + \beta \text{sin}\Omega t)} &= E e^{i \omega t} \Big( J_0(\beta) + \sum_{n=1}^{\infty} J_n(\beta)e^{i n \Omega t} \\ \nonumber  & + \sum_{n=1}^{\infty} (-1)^n J_n(\beta)e^{-i n \Omega t} \Big).
\end{align}
An expansion of the phase modulation term expresses the incident field as a sum of sidebands, each with an amplitude characterized by Bessel functions $J_n(\beta)$ and modulation depth $\beta$. While the EOM has the convenient feature of placing the first order sidebands at $\pm \Omega$, 180$^{\circ}$ out of phase with each other, it has the undesirable feature that deep modulation generates higher order sidebands at frequencies $\pm n\Omega$ with $n>1$ (Fig.~\ref{fig:schematic_opt} blue inset). The presence of these higher order sidebands can reduce the frequency discrimination slope, introduce additional RAM, and limit the dynamic range of the carrier used to probe the desired resonance. 

Here we report a new PDH approach that generates only the three useful tones for frequency discrimination, which are translated into the optical domain with an acousto-optic modulator (AOM). This allows for arbitrary control of each tone strength without bringing other unrelated noise terms from high-order sidebands. This feature also provides a higher signal to noise ratio (SNR), limited by shot noise. Furthermore, EOMs typically generate significant RAM in the context of state-of-the-art optical oscillators, owing to the natural crystal birefringence. It is often desirable to use an active cancellation scheme to prevent RAM from contaminating the amplitude modulation generated by the cavity or atomic and molecular resonances~\cite{Wong1985,Ye1998,Foltynowicz2011,Jaatinen2008,li2012measurement}. Active cancellation schemes that leverage the birefringence of the crystal address only the phase quadrature but not the amplitude imbalance between sidebands. A number of recent research efforts have additionally investigated methods to reduce the EOM's intrinsic RAM~\cite{HallIEEE,Dooley2012,Sathian2013,Zhang2014,Chen2015,Diehl2017,Gillot2022,Bi2019,li2016reduction}. Synthesizing the PDH triplet electronically and translating it into the optical domain by an AOM leads to a lower level of RAM, and provides a natural basis for feedback corrections in the RF domain. These electrical feedback actions are direct and orthogonal, leading to an ideal cavity interrogation beam, limited only by the shotnoise in the guide error signal from RAM PD.
We also note that all components of frequency synthesis, demodulation, and control can be integrated into a compact electronic unit, leading to significant reduction of both the cost and complexity of performing FM spectroscopy. 

General realizations of PDH locking rarely venture into the regime of $\beta > 1$, but recent work with state-of-the-art optical oscillators has identified cases where a suppressed carrier and high modulation tones (generated with $\beta > 1$) can lead to lower frequency drift \cite{Robinson4K} in dielectric coatings, or reduced light-induced effects in crystalline coatings \cite{Jialiang,Dhruv}. 
The use of low optical carrier power has been necessary in achieving the best long term stability of optical oscillators. This is desirable in demonstrating the role of cryogenic crystalline cavities for all-optical timescales~\cite{Milner2019} and searches for novel fundamental physics~\cite{Kennedy2020}. 
In the traditional EOM-based FM approach, it is usually more challenging to reduce and stabilize RAM when $\beta \gg 1$. To make further advances in long-term stability, the independent and arbitrary amplitude control of the carrier and modulation sidebands is therefore a highly desirable feature. We realize this goal using the versatile control offered by our synthetic FM to demonstrate that we can achieve active RAM control at the fractional $10^{-7}$ level relative to the cavity linewidth even with a vanishingly small optical carrier.

\begin{figure}[h!]
    \begin{subfigure}[t]{0.5\textwidth}
        \includegraphics[width=\textwidth]{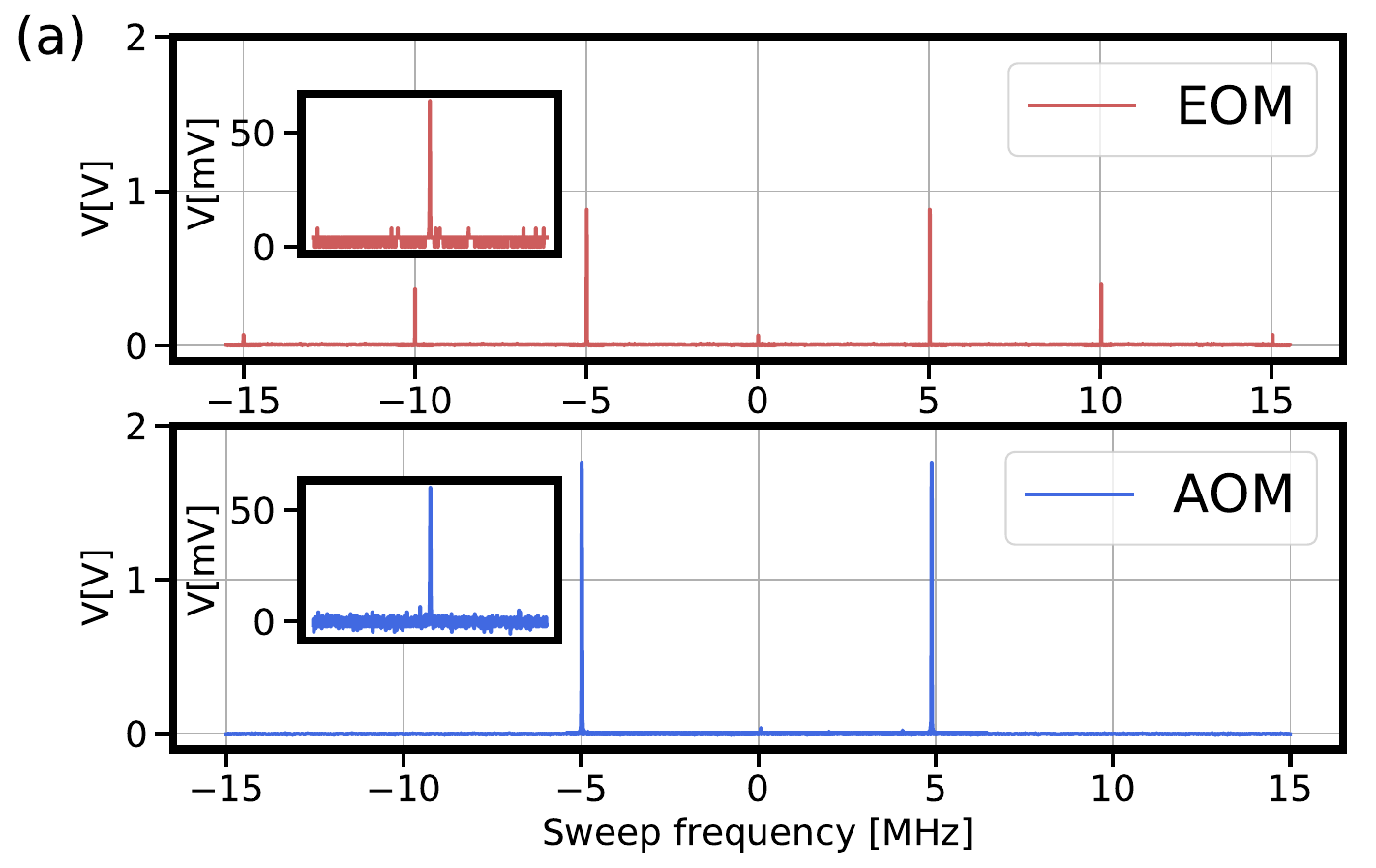}
    \end{subfigure}
    \begin{subfigure}[t]{0.5\textwidth}
        \centering
        \includegraphics[width=\textwidth]{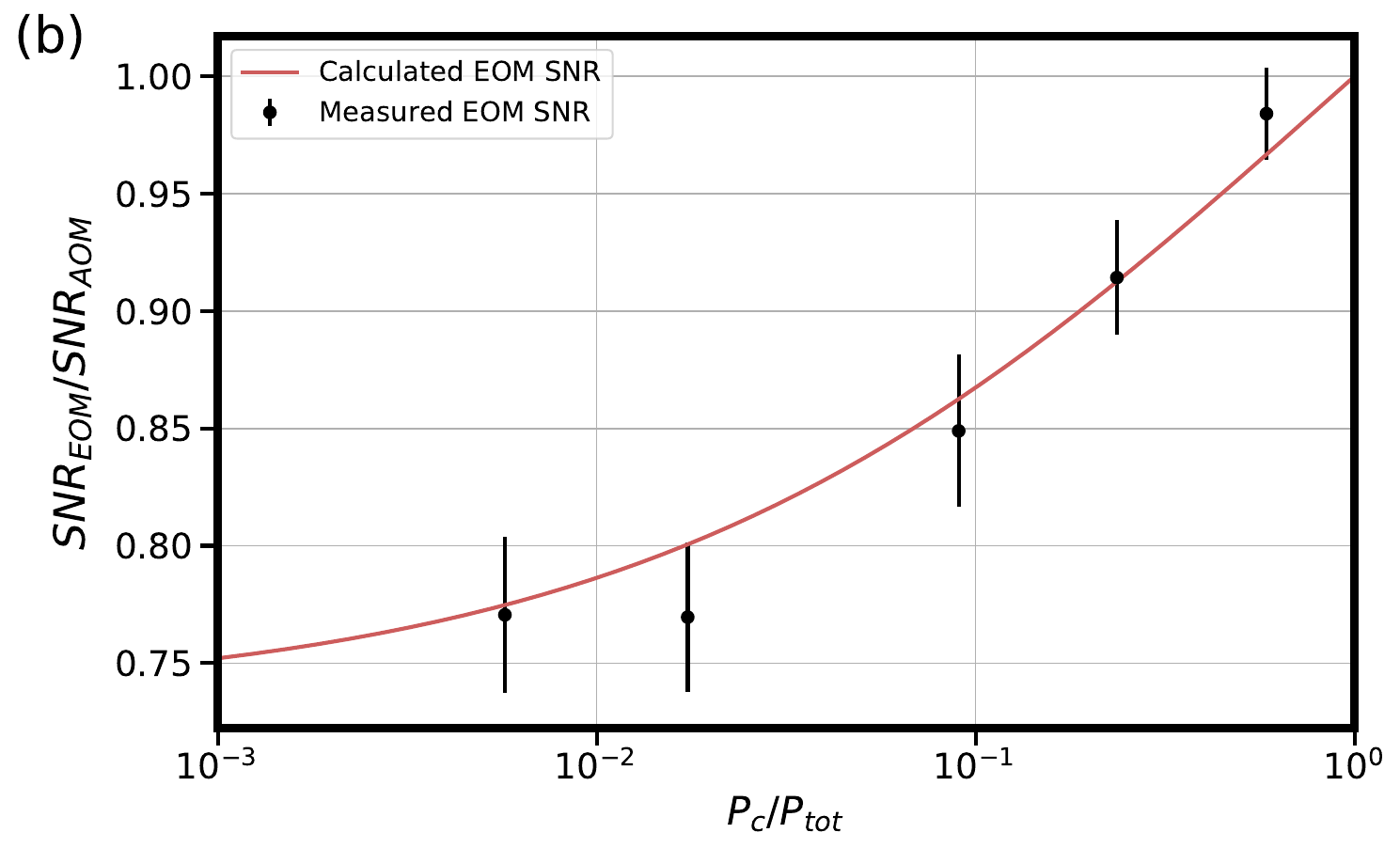}
    \end{subfigure}
        \caption{(a): The cavity transmission of EOM (top) and AOM (bottom) configurations when the $P_c/P_{tot}=0.5\%$ in the case of modulation index of EOM $\beta \approx 2.3$. Each inset plot shows the measured carrier transmission as the modulated laser is swept in frequency.
        (b): Comparison of the shot noise limited signal to noise ratio (SNR) between the EOM and AOM modulation schemes.  The total incident power $P_{tot}$ is fixed, and the fraction of carrier power $\frac{Pc}{P_{tot}}$ is varied. A modulation index of $\beta = 1$ roughly corresponds to $\frac{Pc}{P_{tot}} \approx 0.5 $. The measured results: black dots;  calculation: red line. 
        }
        \label{fig:SNR}
\end{figure}
\section{Methods}
Figure~\ref{fig:schematic_opt} shows the entire experimental setup, including the PDH locking schematic and active RAM servo for both AOM and EOM configurations. In the AOM digital configuration, a carrier $f_0$ and equally spaced modulation sidebands $f_0 \pm \Omega$ (with $f_0 = 80$ MHz and $\Omega=5$ MHz throughout this text) are generated via a Direct Digital Synthesizer (DDS).  Crucially, we adjust the output phase of one of the DDS channels such that one of the modulation sidebands is 180$^\text{o}$ out of phase with the carrier, mimicking the electric field output of a phase modulated EOM. All three tones are referenced to a common 1 GHz clock and a fourth channel on the DDS synthesizes the RF local oscillator (LO) that is used in the demodulation process. The three RF tones drive a fiber coupled AOM so that the first diffracted orders (at different frequencies) are coupled together and propagate in the same fiber. Upon outcoupling through a collimator, they then all share an identical spatial path.

We split the light into two paths - one is incident on a photo detector (PD) used for measurement and active stabilization of RAM, and the other is incident on a detector for PDH locking placed in cavity reflection. Each photodetector has resonant gain at our modulation frequency of 5 Mhz, further suppressing pickup of any high RF signals. In this work the PDH PD is instead used to characterize the out-of-loop (OoL) RAM of the entire system when the laser frequency is tuned sufficiently far away from the cavity resonance and the corresponding baseline fluctuations are measured as the OoL RAM. The cavity used in the experiment is 6 cm long with finesse 497,000 corresponding to a cavity linewidth of 5 kHz. We compare this scheme to traditional EOM-based PDH by replacing the AOM with a LiNbO3 fiber-coupled waveguide EOM (bottom of Fig.~\ref{fig:schematic_opt}), which was used in our previous cavity-locking efforts. All other optical elements remain the same.

One intrinsic advantage of synthesizing the modulation triplet via three separate tones is the absence of higher order sidebands generated when phase modulating an EOM (Fig.~\ref{fig:SNR} (a)). Consequently, there is a higher signal to noise offered in the former case since there are no additional shot noise contributions from the higher modulation harmonics. We illustrate this effect in Fig.~\ref{fig:SNR}(b) where a high modulation index $\beta \approx 2.3$ (corresponding to a small ratio of carrier power $P_c$ over total power $P_{tot}$) is used for the operation for cryogenic silicon cavities with very small long-term frequency drift~\cite{Robinson4K,Milner2019,Dhruv}. In the regime of $P_c / P_{tot} \ll$ 1, the shot noise limited SNR for PDH detection via an EOM is only 75\% of that with the optimal modulation scheme of AOM. 
\begin{figure*}
    \centering
    \includegraphics[width=0.95\linewidth]{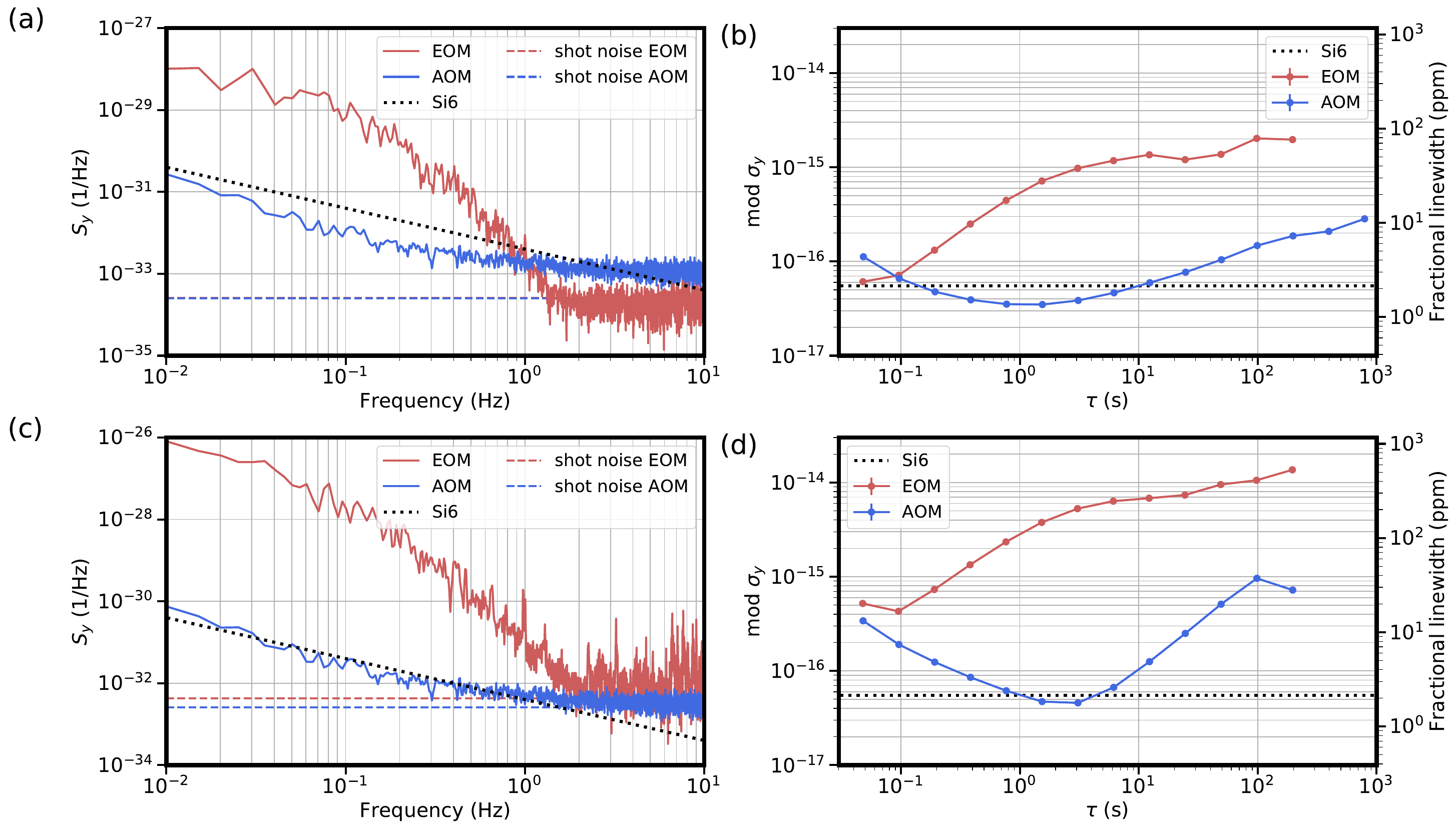}
    \caption{The measurement of the free running RAM, (a) and (c): The power spectral density of RAM-induced frequency noise for high (top, $P_c/P_{tot} = 0.6$) and low (bottom, $P_c/P_{tot} = 0.005$) carrier power ratio. The red and blue dashed lines denote the shot noise limited frequency noise while the black dotted line denotes Si6 noise as measured in \cite{Dhruv}. (b) and (d): RAM-induced fractional frequency instability for the corresponding plots above using the AOM and EOM modulation schemes. Free running RAM for the AOM configuration exhibits greater than an order of magnitude improved performance over EOM at longer integration times in both situations. The right axis plots the data as fractional noise relative to the cavity linewidth.}
        \label{fig:free_run}
\end{figure*}
Next we compare the passive stability of RAM between the EOM and AOM methods (Fig.~\ref{fig:free_run}). In EOM-based PDH, a misalignment of the input field polarization to the principal axes of the modulator leads to a polarization-dependent modulation due to differing refractive indices of the two crystal axes. Thermal variations in the crystal birefringence can then be a significant source of RAM and are often the dominant contribution. The modulation RF field additionally has a spatially inhomogeneous distribution inside the EOM crystal, resulting in another source for RAM.  Conversely, triplet synthesis with an AOM is not subject to the same dispersive effects between the carrier and sidebands. Both methods are sensitive to intra-crystal etalons, which can be mitigated by an angle cut of the crystal faces, but notably the fiber-AOM is found to be relatively insensitive to drifts in the input polarization of the incident light field. The primary source of RAM in the AOM scheme is generated during the triplet synthesis itself. Differential amplitude and phase fluctuations of the synthesizer outputs will appear as noise in the I and Q quadratures at the modulation frequency $\Omega$ but are suppressed through the operation of the two RAM servo loops acting directly on the amplitude and phase quadratures of the two sidebands. Replacing optical issues by a stable electronic system in this manner is an advantageous exchange as the free running RAM can be reduced by careful attention to the RF synthesis, which exhibits a much more robust RAM suppression over a long time scale, as depicted in Fig.~\ref{fig:free_run}. 

State-of-the-art optical oscillators exhibit $<$1$\times 10^{-6}$ (sub-ppm) fractional stabilization to the cavity linewidth, which requires active stabilization of the RAM. In EOM-based PDH, active stabilization can be achieved by temperature controlling the electro-optic crystal and biasing it with a DC voltage that affects the light polarization \cite{Zhang2014,Chen2015,Gillot2022}. Two issues are therefore present with this modulation method: the modulation device itself generates unacceptable RAM, and the actuators for RAM stabilization are not aligned to the in-phase (I) and quadrature (Q) components of the electric field.

\begin{figure*}
    \centering
    \includegraphics[width=0.95\linewidth]{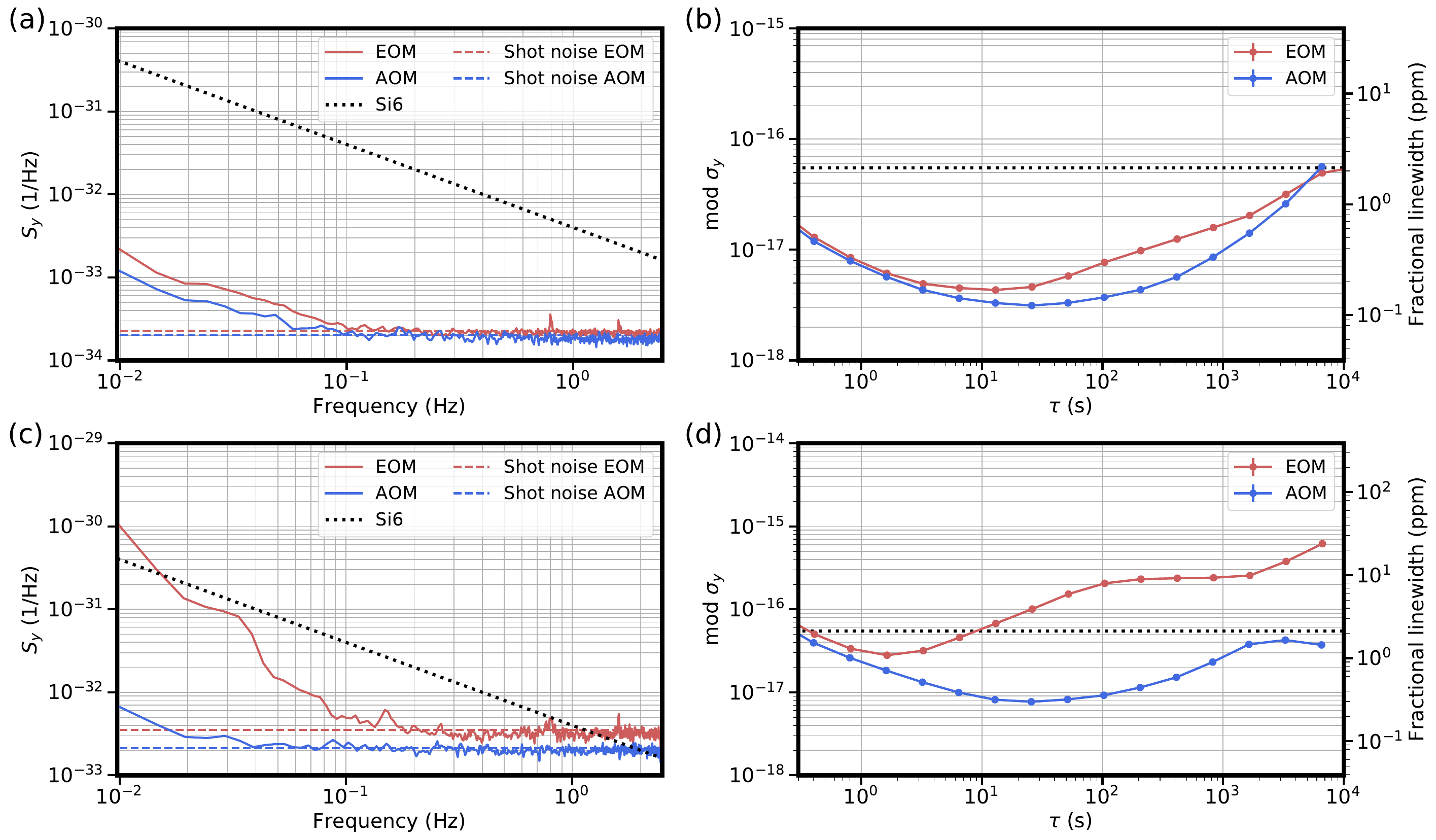}
    \caption{(a) and (c): The power spectral density of RAM-induced frequency noise for high (top, $P_c/P_{tot} = 0.6$) and low (bottom, $P_c/P_{tot} = 0.005$) carrier power ratio. All measurements actively stabilize the RAM to noise detected at the RAM PD. The measured signals are out of loop RAM contributions detected by the PDH PD. The black dotted line presents the current level of noise measured on the 6 cm silicon cavity (see~\cite{Dhruv} for details). Dashed lines denote the shot noise limited frequency noise. (b) and (d): RAM-induced fractional frequency instability for the corresponding plots above using the AOM and EOM modulation schemes.}
        \label{fig:results}
\end{figure*}

Even with the superior passive RAM stability demonstrated by the AOM triplet scheme, we can easily improve the performance further by phase locking three low-phase noise crystal oscillators to each other with an integer frequency synthesis based on a common clock oscillator, frequency division and single sideband modulation techniques. Here we present a unique RAM stabilization scheme to demonstrate that this is not necessary. Specifically, we implement dual quadrature I/Q demodulation of the signal detected on the RAM PD. Rather than tune crystal parameters to affect the output field, we directly feedback to the RF outputs of the DDS. An attenuator represented by a voltage-controlled channel conductance of a junction field effect transistor on the $f_0 + \Omega$ channel output receives a control voltage addressing the error signal of the mixer's I output. Phase noise at $\Omega$ is actively removed by dynamic tuning of the resonance of an inductor-capacitor (represented by two varactor diodes in anti-series configuration) circuit placed at the output of the $f_0 + \Omega$ channel. Orthogonality between the mixer's I and Q outputs is maintained by means of a wire-and-core 3dB quadrature hybrid coupler. The amplitude and phase noise feedback loops are both closed with bandwidths of 120 kHz. The in-loop performance of these stabilization loops is well below the detection shot noise, demonstrating that high suppression of the RAM signal is easily achievable with this orthogonalized and simple controller scheme. A comparison of active RAM stabilization between the AOM and EOM schemes is shown in Fig. \ref{fig:results} and we discuss the implications below.

\section{Results}
We benchmark performance of the AOM modulation scheme against EOM-based PDH by using the optical setup illustrated in Fig.~\ref{fig:schematic_opt} to compare free running and out-of-loop (OoL) RAM performances. This figure of merit will ultimately determine the contribution of RAM to the cavity-stabilized laser frequency instability. We compare the two methods first in the regime of conventional PDH, where the modulation index is set to $\beta \approx 1$ and the PDH error signal is maximized by setting the sideband power to nearly half of the carrier power. We set the carrier power $P_c$ to be equivalent in both setups (in this case, $P_c = 3.3$ $\mu$W and $P_{tot} = 5.5$ $\mu$W, or $\beta=1.1$) such that the corresponding frequency discrimination slopes for both approaches are nearly identical.

Notably, we observe a significant improvement in free running RAM control with the AOM when compared to the EOM (Fig.~\ref{fig:free_run}). Simply adopting the presented scheme without feedback enables control of the optical oscillator to the high $10^{-17}$, a significant standard when compared to state-of-the-art results. High-ppm level stabilization to the cavity linewidth is obtainable using synthetic FM from the 0.1-100 s timescale relevant for many metrology experiments. The lower AOM RAM performance with feedback disabled is even more pronounced when greatly reducing the carrier power relative to the sideband power ($P_c/P_{tot} = 0.005$, Fig.~\ref{fig:free_run}(b)). Enabling RAM stabilization, the resulting OoL RAM signals as measured on the PDH PD are compared in Fig.~\ref{fig:results}, displayed in (a) power spectral density and (b) modified Allan deviation. The fractional frequency noise at bandwidth above 0.1 Hz is photon shot noise limited to the same value in both methods. Even in this conventional PDH regime the AOM-based triplet synthesis demonstrates a lower instability at long integration times, indicating a slightly improved performance at lower Fourier frequencies. Electronic noise contributions to RAM stabilization (detection noise, finite servo gain) are well below the measured OoL traces, and the measured instability is attributed to noise on the optical field. The Allan deviation data from the two schemes converge to a similar level of noise at averaging times of $\tau > 4000$ s as shown in Fig.~\ref{fig:results}(b), suggesting the presence of path length fluctuations that are common to both setups between the in-loop and out-of-loop detectors. Basically, for $\beta \approx 1$, the two schemes perform at a similar level. The AOM triplet offers only a slight advantage in performance, but a significant advantage in experimental construction and implementation.  

For cryogenic silicon cavities operated under 4 K and 16 K~\cite{Robinson4K,Dhruv,Milner2019}, long-term cavity instability is observed to improve with significantly reduced optical carrier power that is coupled to the inside of the cavity. Thus, we now consider the regime of $P_c / P_{tot} \ll 1$. The carrier power is reduced to $P_c = 100$ nW, and we again keep the total optical power is identical for both setups, $P_{tot} = 20$ $\mu$W (Fig.~\ref{fig:results}(c,d)). Note that the signal-to-noise ratio obtained from the AOM modulation is now 22$\%$ higher than that received via EOM-based PDH, and the detected RF signal can be arbitrarily increased with the independent control of sideband amplitude. At high Fourier frequencies where the fractional frequency noise power spectral density is consistent with the photon shot noise limit, both methods show similar instability that is now 10 dB higher owing to the reduced carrier power. However, performance at the low Fourier frequencies shows a dramatic advantage offered by the AOM triplet scheme, where the OoL RAM limitation is still well below the measured cavity noise. This extreme limit of low $P_c$ is exactly the regime where one needs to operate cryogenic silicon cavities in order to enjoy the unparalleled short and long term stability~\cite{Oelker2019}. 

We again verify that the OoL RAM is not limited by electronic contributions. Direct orthogonalized phase and amplitude control to the frequency triplet synthesized with the AOM method offers order of magnitude improved performance over the EOM scheme, where the high modulation index has generated large amplitude modulation tones at frequencies $\pm 2\Omega$ and $\pm 3\Omega$. The significant improvement of the AOM scheme indicates that there is a fundamental difference between the ways that the two methods respond to any out-of-loop noise process. A prominent example is the presence of a parasitic etalon that could be located in the out-of-loop path probed by the PDH detector. Such an effect will generate an amplitude and phase imbalance between the modulation sidebands. Noise originating from the etalon instability will be accentuated in the presence of higher order modulation harmonics. Even with the weak carrier power, RAM is still suppressed to the 0.3 ppm level in the AOM configuration, and such performance demonstrates the potential for long term laser frequency instablity well below $1\times 10^{-16}$, especially when higher cavity finesse is employed. 

\section{Conclusion}
In this article, we present a new method of modulation, in which the PDH frequency triplet is digitally synthesized and frequency-shifted into an optical field with FM by means of an AOM, working with the emitted field of a stable singlemode laser. This new configuration eliminates higher order sidebands generated via phase modulation and consequently offers a higher shot noise limited SNR. Direct and independent feedback to amplitude and phase of the modulated electric field are performed in the RF domain rather than the optical. The versatility and robustness of this scheme present a economic way to control the carrier power independent of the sideband amplitude, allowing one to couple a flexible amount of optimal optical power into the optical cavity. 

We note that this modulation scheme can be extended to experiments requiring a dynamic sweep of the FM tone. Varying the modulation frequency in an EOM results in an altered modulation depth but no such effect occurs in our triplet synthesis scheme. Additionally, we want to point out that that retroreflecting the free space signal before the cavity and allowing it to repass the fiber AOM will heterodyne beatnote at a frequency of $2f_0$, which can be used for path length stabilization. A similar approach enacted in a PDH setup employing an EOM would require a separate AOM for this purpose and we can effectively combine two stabilization loops with a single actuator. Dedicated research efforts have sought to improve the innate RAM associated with EOMs but here we demonstrate a significant improvement in the passive stability of RAM when using the AOM triplet. With active RAM cancellation we also surpass the stability obtained with an EOM, especially in the regime of low carrier power. This is a direct path forward to achieving a lower frequency drift in cryogenic crystalline cavities employing dielectric coatings and reduction of light-induced noise in those with crystalline coatings. The compact and robust nature of this approach also makes PDH and other forms of FM spectroscopy more easily accessible in a wide range of experiments.   
\\
\\
\noindent\textbf{Acknowledgments} The authors thank T. Schibli and J. Sherman for careful reading of the manuscript and C.-Y. Ma for insightful discussions.
\\ \\
\noindent\textbf{Funding} This work is supported by NIST, DARPA, AFRL, and the National Science Foundation Physics Frontier Center (NSF PHY-1734006).
\\ \\
\noindent\textbf{Disclosures} The authors declare no conflicts of interest.
\\ \\
\noindent\textbf{Data Availability} Data underlying the results presented in this paper are not publicly available at this time but may be obtained from the authors upon reasonable request.
\\ \\

\bibliography{refs}

\end{document}